\documentclass[]{aa} 
\usepackage{graphicx,  natbib, epsfig, amsmath, amssymb, mathrsfs, txfonts,bbm,xspace,alltt,textcomp,gensymb}
\usepackage{soul}
\bibpunct{(}{)}{;}{a}{}{,}
\usepackage[colorlinks=true,citecolor=blue]{hyperref}

\setstcolor{magenta}

\begin{document}

\title{Calibration of the island effect: Experimental validation of\\closed-loop focal plane wavefront control on Subaru/SCExAO} 
\titlerunning{Calibration of the island effect}

\author{
M. N'Diaye \inst{1} \and 
F. Martinache \inst{1} \and
N. Jovanovic\inst{2,3} \and
J. Lozi\inst{2} \and
O. Guyon\inst{2,4,5,6} \and
B. Norris\inst{7} \and
A. Ceau \inst{1} \and 
D. Mary\inst{1}
}
\institute{Universit\'e C\^ote d\'{}Azur, Observatoire de la C\^ote d\'{}Azur, CNRS, Laboratoire Lagrange, Bd de l\'{}Observatoire, CS 34229, 06304 Nice cedex 4, France\\
\email{\href{mailto:mamadou.ndiaye@oca.eu}{mamadou.ndiaye@oca.eu}} 
\and 
National Astronomical Observatory of Japan, Subaru Telescope, National Institutes of Natural Sciences, Hilo, HI 96720, USA
\and
Department of Physics and Astronomy, Macquarie University, 2109 Sydney, Australia
\and
Astrobiology Center, National Institutes of Natural Sciences, 2-21-1 Osawa, Mitaka, Tokyo, Japan
\and
Steward Observatory, University of Arizona, Tucson, 933 N Cherry Ave, Tucson, AZ 85721, USA
\and
College of Optical Science, University of Arizona, 1630 E University Blvd, Tucson, AZ 85719, USA
\and 
Sydney Institute for Astronomy (SIfA), School of Physics, University of Sydney, NSW 2006, Australia.
} 

\date{}

\abstract
{Island effect (IE) aberrations are induced by differential pistons, tips, and tilts between neighboring pupil segments on ground-based telescopes, which severely limit the observations of circumstellar environments on the recently deployed exoplanet imagers (e.g., VLT/SPHERE, Gemini/GPI, Subaru/SCExAO) during the best observing conditions. Caused by air temperature gradients at the level of the telescope spiders, these aberrations were recently diagnosed with success on VLT/SPHERE, but so far no complete calibration has been performed to overcome this issue.}
{We propose closed-loop focal plane wavefront control based on the asymmetric Fourier pupil wavefront sensor (APF-WFS) to calibrate these aberrations and improve the image {quality} of exoplanet high-contrast instruments in the presence of the IE.}
{Assuming the archetypal four-quadrant aperture geometry in 8\,m class telescopes, we describe these aberrations as a sum of the independent modes of piston, tip, and tilt that are distributed in each quadrant of the telescope pupil. We calibrate these modes with the APF-WFS before introducing our wavefront control for closed-loop operation. We perform numerical simulations and then experimental tests on a real system using Subaru/SCExAO to validate our control loop in the laboratory and on-sky.}
{Closed-loop operation with the APF-WFS enables the compensation for the IE in simulations and in the laboratory for the small aberration regime. Based on a calibration in the near infrared, we observe an improvement of the image quality in the visible range on the SCExAO/VAMPIRES module with a relative increase in the image Strehl ratio of 37\%.}
{Our first IE calibration paves the way for maximizing the science operations of the current exoplanet imagers. Such an approach and its results prove also very promising in light of the Extremely Large Telescopes (ELTs) and the presence of similar artifacts with their complex aperture geometry.}

\keywords{
    instrumentation: high angular resolution --  
    instrumentation: adaptive optics --
    techniques: high angular resolution  -- 
    telescopes --
    methods: data analysis} 
\maketitle

\section{Introduction}
High-contrast observations have been unveiling planets and debris disks around nearby stars with images and spectra of unprecedented sensitivity and inner working angle \citep[e.g.,][]{Macintosh2014,Chilcote2015,Vigan2015,Currie2017,Samland2017,Rajan2017,Chauvin2017arXiv} thanks to the recent deployment of extreme adaptive optics (ExAO) facilities with coronagraphic capabilities on 8\,m class ground-based telescopes, such as the VLT/SPHERE, Gemini/GPI, or Subaru/SCExAO \citep{Beuzit2008,Macintosh2008,Jovanovic2015}. Such data provide photometric, spectroscopic, polarimetric, and astrometric information on the observed exoplanets and disks to constrain their nature \citep{Oppenheimer2009,Traub2010,Bowler2016} and allow the community to advance in our understanding of the formation, architecture, and dynamics of circumstellar environments \citep[e.g.,][]{Nesvold2015,Boccaletti2015,Dong2017}. So far, these groundbreaking exoplanet imagers have been able to detect substellar mass companions down to warm gas giant exoplanets with contrast ratio up to $10^5$-$10^6$ at separations as low as 0.2-0.3 arcsec in the near infrared. The observation of lighter or colder objects orbiting their host star with larger contrast ratios at smaller separations is, however, currently limited by several hurdles.

The main limiting factor lies in the presence of the noncommon path aberrations (NCPA) between the ExAO sensing path and the science path in the exoplanet imagers. The low-order modes such as pointing errors or focus drift induce misalignments of the star image on the coronagraphic mask of the instrument, leading to a degradation in starlight attenuation. This issue for exoplanet imagers is mainly addressed with dedicated sensors or specific image processing algorithms to achieve precise pointing of the observed star image on the coronagraph \citep[][]{Baudoz2010,Savransky2013,Huby2015,Singh2015}. Higher-order modes, which evolve slowly with time, produce static and quasi-static speckles that swamp the signal from the faint exoplanets in the coronagraphic images. The origin of these speckles was suspected at the time of instrument construction \citep[e.g.,][]{Fusco2006} and preliminary strategies using techniques such as phase diversity or a Mach-Zehnder interferometer were adopted to address these issues \citep{Sauvage2007,Wallace2010}. As these aberrations notably alter the quality of the science data, further research on NCPA metrology methods have been pursued to calibrate these errors down to nanometric accuracy and different solutions have emerged over the past few years to overcome this issue \citep[e.g.,][]{Galicher2010,Martinache2013,N'Diaye2013a,Paul2013}. Some of these solutions are currently being implemented in the current exoplanet imaging facilities or envisioned in preparation for the possible forthcoming upgrades of these instruments to control NCPA during science operations \citep{Jovanovic2015,Fusco2016}. 

Unfortunately, other limiting factors were unexpected at the time of instrument development, one of the most infamous being the island effect (IE).\footnote{We believe that this term was first introduced by Noah Schwartz at the AO4ELT5 conference in Tenerife, Spain (see \url{http://www.iac.es/congreso/AO4ELT5/pages/scientific-programme.php}).} 
These phase discontinuities within the telescope pupil across large gaps (usually spiders) have multiple causes. First, they can be strongly excited by thermal effects at the spiders, in what has been referred to as the low wind effect \citep[LWE, ][]{Sauvage2016}. LWE occurs if radiative cooling can drive large temperature gradients when wind-driven convective or conductive thermal equalization is weak. Evolving at temporal frequencies below 1\,Hz, LWE is observed under low outdoor wind speed conditions, for example, wind speeds below 3\,m/s at 30\,m below the platform at the VLT. Second, they are often poorly sensed by conventional wavefront sensors, as spatial coherence across large gaps is usually either lost or poorly used for sensing. For example, a Shack-Hartmann wavefront sensor (WFS) with small subapertures does not coherently mix much light from one "island" to the next. This effect can be generalized to other WFSs as well, and is ultimately due to a loss of spatial coherence across large spatial spans in the pupil.

This spurious effect can be described as a combination of differential piston, tip, and tilt between the light in the quadrants of the telescope pupil. For the image of an unresolved star, such aberrations lead to the degradation of the star point spread function (PSF) by mostly splitting up its first bright diffraction ring, and even its core for large aberrations, into three or four lobes, hence preventing coronagraphs from removing starlight at their optimal capacity. The impact of these aberrations is disastrous for science operations as they occur during the best-suited observing conditions for exoplanet imaging. A significant loss of observation time has been observed on high-contrast instruments, up to 20\% in the case of SPHERE \citep{Sauvage2016}. This effect is not detected by the ExAO wavefront sensors as IE appears in the null space of the sensor response. On SPHERE, the Shack-Hartmann wavefront sensor  sees the effect very partially \citep{Sauvage2016}. On SCExAO, the pyramid wavefront sensor (PyWFS) also presents a poor response to the IE even though such a response slightly improves with small modulation, good Adaptive Optics (AO) correction, and under good seeing conditions. Because of its peculiar nature, this effect represents a challenge to overcome in terms of calibration.

In the exoplanet imaging context, the IE was first successfully diagnosed on SPHERE using ZELDA (which stands for Zernike sensor for Extremely Low-level Differential Aberrations), a Zernike wavefront sensor to calibrate residual aberrations in coronagraphic instruments \citep{N'Diaye2013a,N'Diaye2016b,Sauvage2016}. In parallel to this concept, other sensors for NCPA calibration have been developed and implemented in analogous instruments. Among them is the Asymmetric Pupil Fourier Wavefront Sensor \citep[APF-WFS,][]{Martinache2013}, a focal plane wavefront sensor with a prototype that is installed on SCExAO. Phase diversity methods are also emerging to accurately estimate these pupil phase discontinuities \citep{Lamb2017accepted}. 

While metrology solutions have started to emerge, no complete calibration of this effect has yet been performed on an ExAO instrument to the best of our knowledge. In this paper, we propose to demonstrate the calibration of the IE for small phase errors with APF-WFS and validate the corresponding control loop on SCExAO. We recall the principle of this focal plane wavefront sensor and detail the control loop operations using this sensor. Experimental results are presented with our concept using SCExAO under in-lab tests and on-sky observations to validate the principle. We finally derive conclusions and discuss their implications for further exoplanet high-contrast observations.

\section{Wavefront control for the island effect}\label{sec:WFC} 
\subsection{Principle and formalism of the APF-WFS}
\begin{figure}[!ht]
\centering
\includegraphics[width=0.5\textwidth]{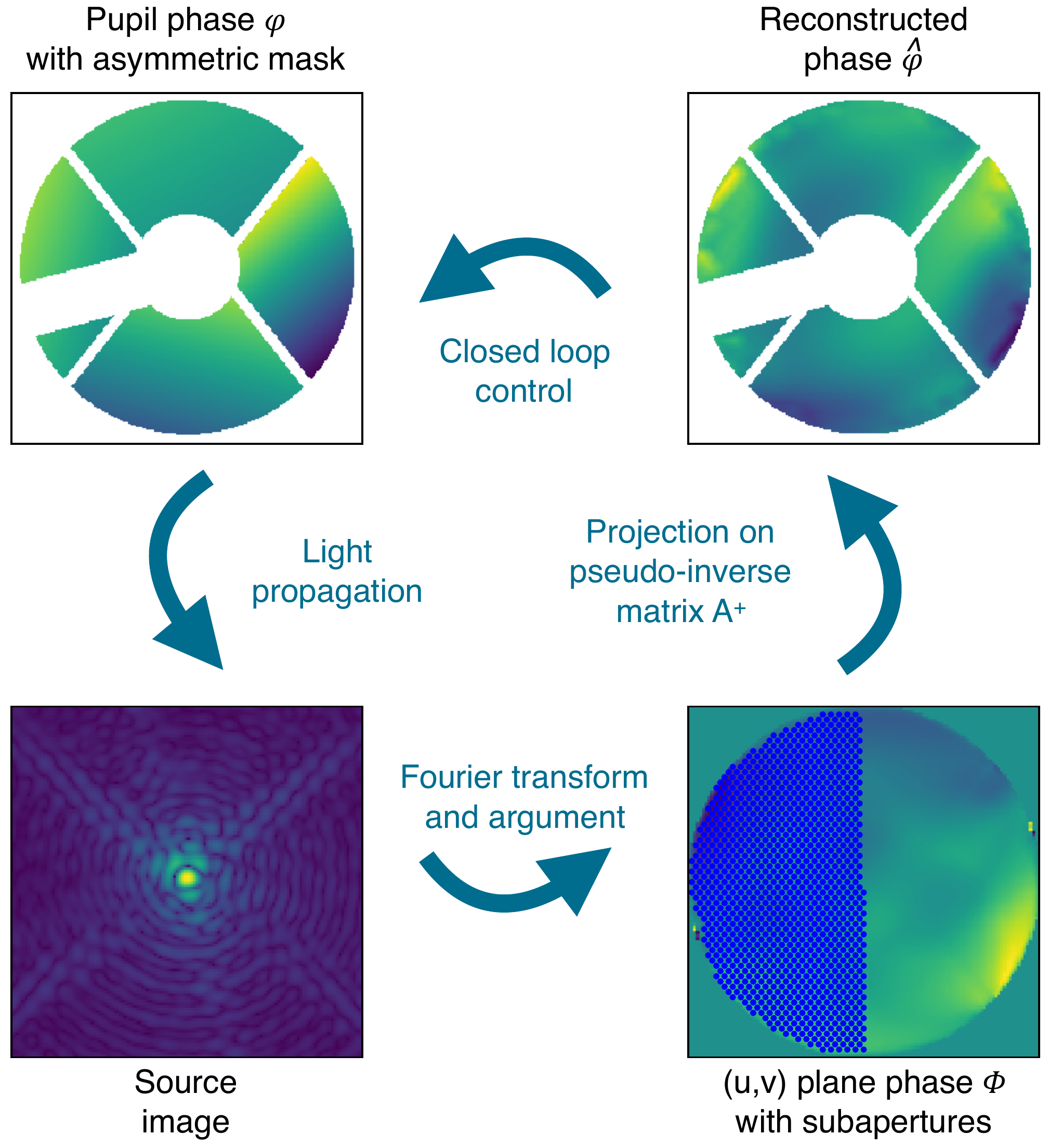}
\caption{Scheme of the phase reconstruction process with APF-WFS: phase map $\varphi$ within the asymmetric pupil mask (top left), source image on the camera (bottom left), phase $\Phi$ of the Fourier-plane signature in the (u,v) plane overlapped with half of the subaperture discretization (bottom right), and the reconstructed pupil phase map $\hat{\varphi}$ from the pseudo-inverse matrix $\mathbf{A}^{+}$ (top right).}
\label{fig:scheme}
\end{figure}

The APF-WFS is based on the Fourier analysis of the image for an observed unresolved star to retrieve the wavefront errors in the system \citep{Martinache2013,Martinache2016}. Figure \ref{fig:scheme} shows a schematic view of the APF-WFS principle. 

The shape of a PSF is the result of starlight diffraction due to the telescope aperture and on-sky turbulence for a ground-based telescope. The star image structure can equivalently be described from an interferometric standpoint by considering the telescope pupil as a combination of subapertures by means of pupil discretization. The PSF therefore represents the result of optical interferences between all the subapertures. Performing a Fourier transform of the star image thus allows one to analyze the PSF in the Fourier plane. From an interferometric point of view, this is the (u,v)-plane and the coverage is a result of the baselines that are formed between all the interfering subapertures. This operation leads to a relation between the image plane intensity and its Fourier transform. 

The phases in the pupil plane $\varphi$, and in the Fourier space $\Phi$, are linearly related in the small aberration regime ($\varphi \ll$1\,rad, corresponding to 250\,nm root mean square (RMS) wavefront errors in $H$-band at 1.6\,$\mu$m). Using pupil discretization and in the absence of companions or disks in the astrophysical scene, this relation can be expressed as 
\begin{equation}
    \Phi = \mathbf{A}\,\varphi\,,
    \label{eq:phase_relation}
\end{equation}
in which $\mathbf{A}$ denotes the transfer matrix between the two types of phase. By inverting this relation, we have access to the wavefront error but are left with the problem of its sign ambiguity. To lift this degeneracy, \citet{Martinache2013} introduced an asymmetry in the pupil by means of a mask, leading to an unambiguous determination of the wavefront error.

As the matrix $\mathbf{A}$ is rectangular, we produce its pseudo-inverse $\mathbf{A^{+}}$, with $\mathbf{A^{+}=(A^T\,A)^{-1}\,A^T}$. Such an operation can be performed by implementing singular value decomposition (SVD) of the original matrix. The corresponding singular values spread over several orders of magnitude and thus, the smallest values have a lower impact on the phase reconstruction and can hence be set to zero without the loss of too much information. By denoting $k$ the number of first ordered singular values that are left non-nulled, the corresponding pseudo-inverse matrix $\mathbf{A_{k}^{+}}$ leads us to an estimate of the wavefront error $\hat{\varphi}$, given by
\begin{equation}
    \hat{\varphi} = \mathbf{A_{k}^{+}}\,\Phi\,.
    \label{eq:inverse_relation}
\end{equation}

\subsection{Calibration of the island effect}
The IE characterization requires the definition of an adequate mode basis for modal closed-loop correction. Most of the 8\,m class monolithic telescopes present a pupil that splits up into four quadrants with the shadow of the secondary mirror and support struts on the primary mirror. Under this pupil geometry, the IE can be visualized as the combination of piston, tip, and tilt in each telescope pupil quadrant, that is, the first three Zernike modes over four quadrants. 

\begin{figure}[!ht]
\centering
\includegraphics[width = 0.5\textwidth]{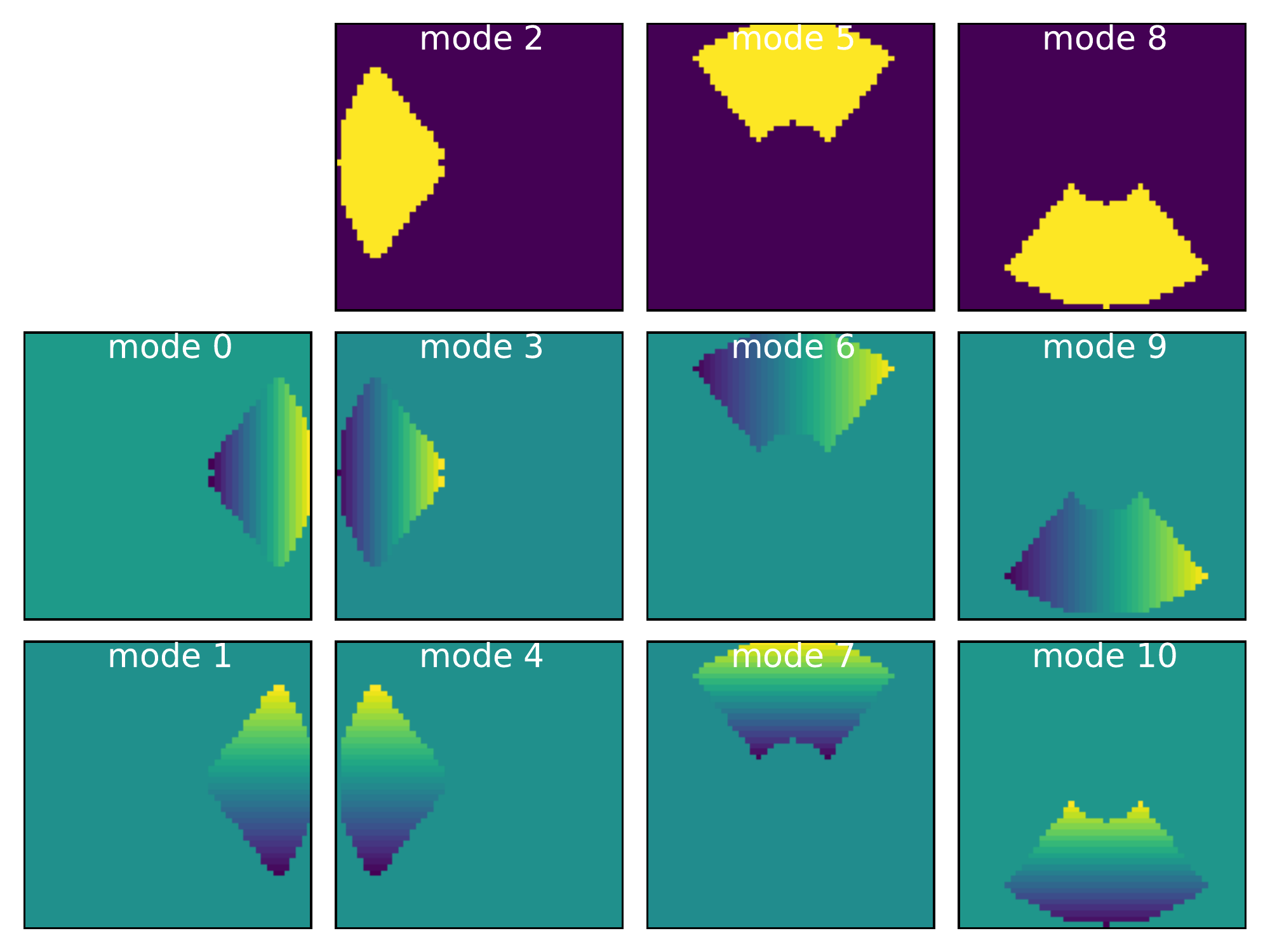}
\caption{Panels of the 11 modes for the DM to control the island effect. Each panel column corresponds to a pupil quadrant and panel rows correspond to piston, tip, and tilt starting from the top.}
\label{fig:11modes_DM}
\end{figure}

These aberrations form a natural set of 12 modes that can be represented in a vector form and gathered in a matrix. Computing the rank for this matrix leads us to a value of 11 instead of 12, showing some redundancy between the modes. We recover an 11-rank matrix by removing one of the four piston modes to form an 11-mode matrix (see Figure \ref{fig:11modes_DM}). Such an operation is useful to ensure the stability of a wavefront control algorithm. In this work, we express the IE modes in RMS wavefront error units, but we can alternatively give them in peak-to-valley (PV) amplitudes since the IE represents a sum of basic modes. For reference, we provide the correspondence between RMS and PV amplitudes for these errors with the Subaru pupil in Table \ref{tab:coeffs_PV_RMS}.

\begin{table}[!ht]
    \centering
    \caption{Peak-to-valley (PV) amplitude for one wavefront error RMS unit for each of the IE modes with the Subaru telescope pupil.}
    \begin{tabular}{ccccccc}
    \hline\hline
    modes & 0, 3 & 1, 4 & 2 & 5, 8 & 6, 9 & 7, 10\\
    \hline
    PV for 1 RMS & 11.01 & 11.45 & 2.70 & 2.39 & 9.66 & 8.47\\
    \hline
    \end{tabular}
    \label{tab:coeffs_PV_RMS}
\end{table}

\begin{figure}[!ht]
\centering
\includegraphics[width=0.5\textwidth]{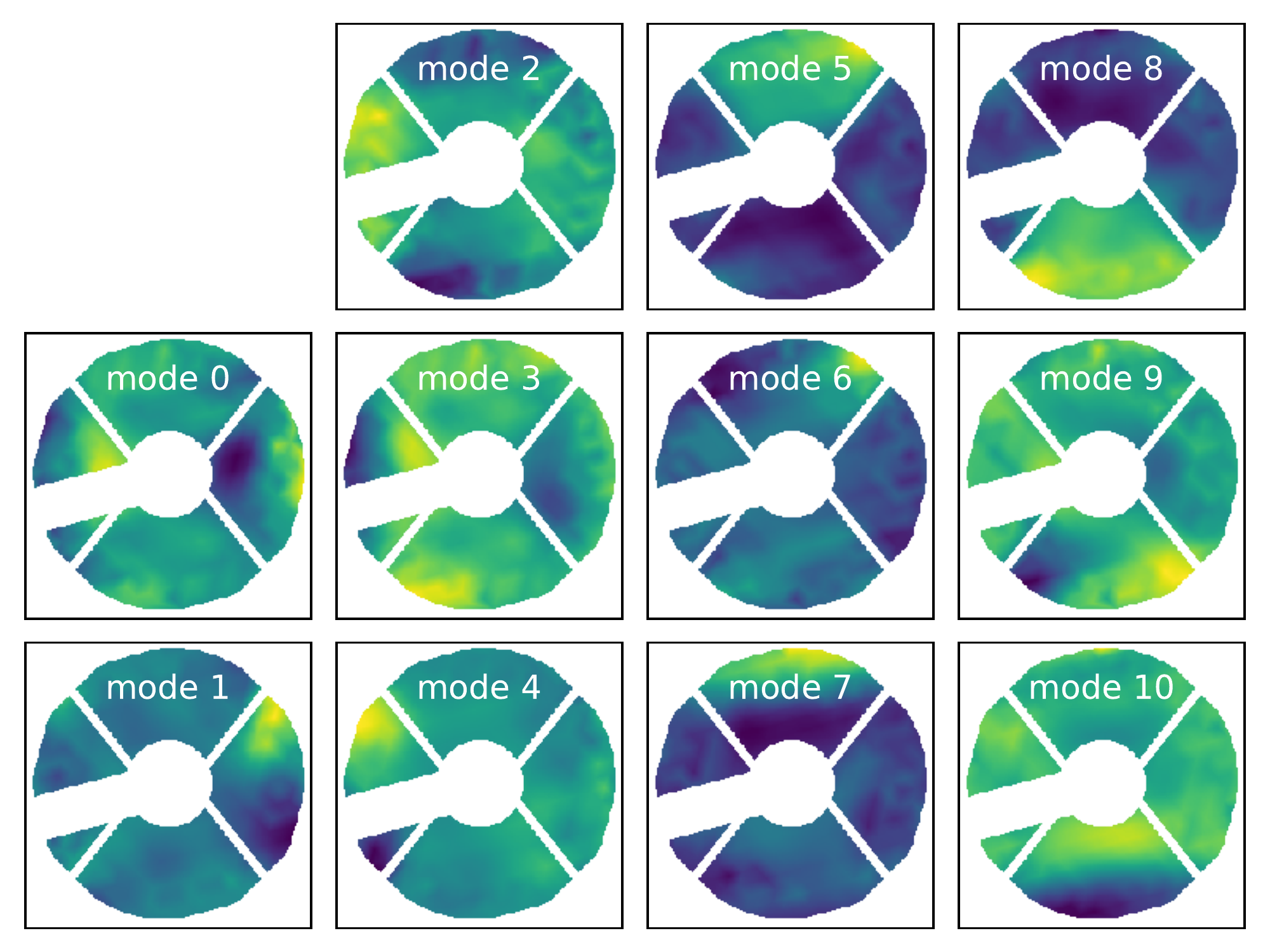}
\caption{Reconstruction of the 11 IE modes with a wavefront error of 100\,nm RMS, using 100 out of the 291 modes that are kept in the model. The modes are represented in the presence of the asymmetric pupil mask. From left to right: pupil quadrants right, left, top, and bottom. From top to bottom: piston, tip, and tilt.}
\label{fig:11modes_response}
\end{figure}

Assuming the Subaru Telescope aperture for the pupil geometry, a closed-loop system with a deformable mirror (DM) and an APF-WFS with observation at the wavelength $\lambda$, we acquire the response of each of these 11 modes and assemble them into a response matrix $\mathbf{L}$: we excite each mode on the DM, produce the respective PSF, retrieve the corresponding phase response from the wavefront sensing analysis, and gather the responses from the 11 modes in a matrix form. For illustration, we work in $H$ band at $\lambda$=1.6\,$\mu$m with the parameters of the APF-WFS that are used on SCExAO \citep{Martinache2016}: a discrete model containing 292 subapertures that are spaced at 0.39\,m across the 7.92\,m telescope pupil and a mask with an arm that is oriented with a position angle of 165$\degree$ in the clockwise direction in the pupil plane and presents a thickness of $13\%$ of the pupil diameter. By selecting one subaperture as a reference, we obtain a total number of 291 singular modes for this model. Figure \ref{fig:11modes_response} displays the reconstructed phase for the 11 modes and a wavefront error of 100\,nm RMS, showing overall satisfactory consistency with the modes introduced by the DM and represented in Figure \ref{fig:11modes_DM}. Discrepancies (in particular modes 0 and 3) would require an updated discrete model of the pupil. We are currently evaluating this option. We, however, decided to keep the model described by \citet{Martinache2016} for this work.

\subsection{Closed-loop operation}\label{subsec:closed-loop}
Our goal now is to compensate for the IE in a closed-loop operation based on our calibration. 
A PSF is acquired in the presence of the asymmetric pupil mask and analyzed with the wavefront sensing algorithm to extract the respective phase $\varphi$. This instant wavefront is a linear combination of the 11 modes with appropriate coefficients $\alpha$,
\begin{equation}
    \varphi = \mathbf{L}\,\alpha. 
    \label{eq:alpha}
\end{equation}
The coefficients $\alpha$ can be estimated by minimizing $||\varphi - \mathbf{L}\alpha ||^2$.
Resolving this least-square problem provides us coefficients $\hat{\alpha}$ that are associated with the 11 modes to produce a map to apply to the DM and perform modal control:
\begin{equation}
    \hat{\alpha} = \mathbf{L}^{+}\,\varphi\,,
\label{eq:hat_alpha_result}
\end{equation}
with $\mathbf{L^{+}=(L^T\,L)^{-1}\,L^T}$. In a real-life system, a control loop gain smaller than $1$ is used to ensure its convergence towards a null wavefront error with the control algorithm.

\begin{figure}[!ht]
\centering
\includegraphics[width=0.5\textwidth]{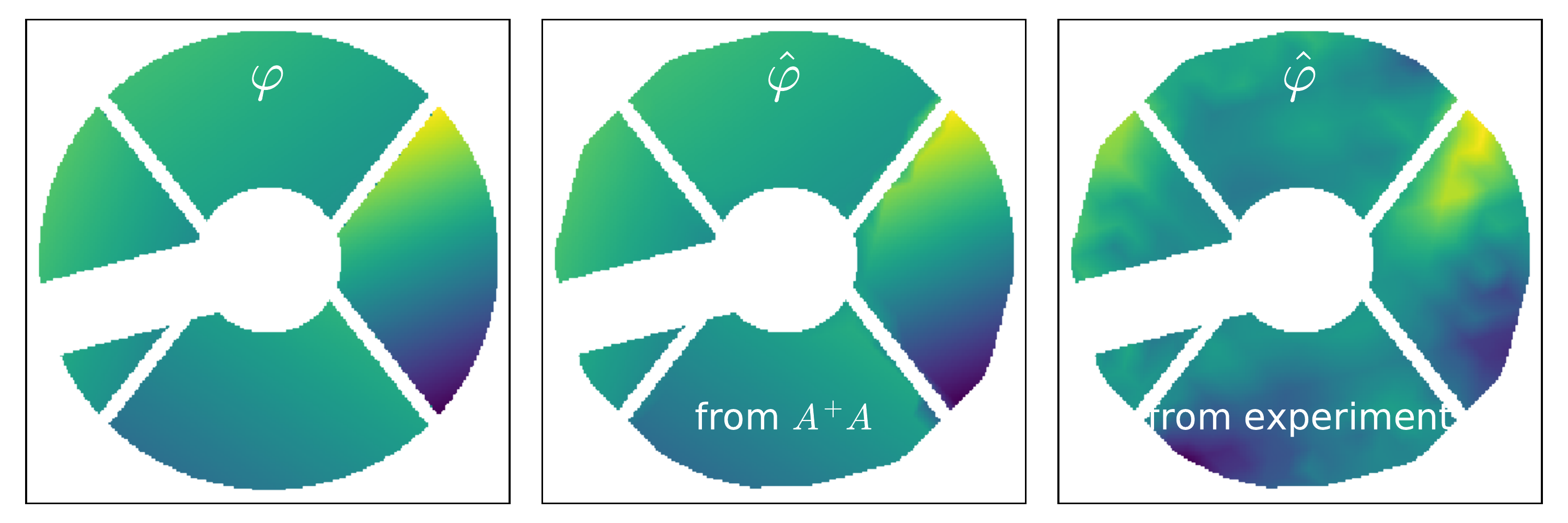}
\caption{From left: initial IE map $\varphi$, reconstructed phase map $\hat{\varphi}$ from the linear model with $\mathbf{A}^{+}\,\mathbf{A}$ and no cutoff, and reconstructed phase map $\hat{\varphi}$ from the experimental process with the linear model and a cutoff of singular modes $k=100$.}
\label{fig:phase_comparison}
\end{figure}

At this point, it is interesting to compare the IE maps resulting from this experimental process and from the product $\mathbf{A}^{+}\mathbf{A}$ from the combination of Eqs. (\ref{eq:phase_relation}) and (\ref{eq:inverse_relation}) to determine the quality of the map reconstruction. We first generate a phase map with a linear combination of the 11 modes with the asymmetric pupil mask, using the coefficients given in Table \ref{tab:coeffs_simu}. The total wavefront error for the resulting phase map is 136\,nm RMS. Our simulations are made in the absence of noise (no shot noise or readout noise). Figure \ref{fig:phase_comparison} shows the initial map in the left panel and the maps resulting from both reconstruction methods in the middle and right panels. For the experimental case, we use a cutoff $k=100$ for the singular modes of the model, justifying this selection hereafter. The reconstructed map in the right panel from the experiment through image processing is visually similar to the ideal reconstruction from the product $\mathbf{A}^{+}\mathbf{A}$ in the middle panel. From a quantitative point of view, the respective wavefront errors inside the pupil are also very close, with 133 and 132\,nm RMS. After computation of the map difference and its wavefront error dispersion inside the pupil, we find a reconstruction error of 74\,nm RMS between both maps, coming from a residual global tip/tilt and the high spatial frequency errors due to the discontinuities at the pupil edge. These results confirm the high quality of our wavefront reconstitution with respect to the theoretical case for the IE modes.

\begin{table}[!ht]
    \centering
    \caption{Amplitude of the modes in wavefront errors to produce the arbitrary IE map used in our simulations, assuming observation in $H$ band ($\lambda=$1.6\,$\mu$m).}
    \begin{tabular}{ccccccc}
    \hline\hline
    modes & 0 & 1 & 2 & 3 & 4 & 5\\
    \hline
    RMS (nm) & -26 & 100 & 30 & -26 & 20 & 34\\
    \hline\hline
    modes & 6 & 7 & 8 & 9 & 10 & \\
    \hline
    RMS (nm) & -16 & 18 & -4 & 24 & 30 & \\ 
    \hline
    \end{tabular}
    \label{tab:coeffs_simu}
\end{table}

\begin{figure}[!ht]
\centering
\includegraphics[width=0.5\textwidth]{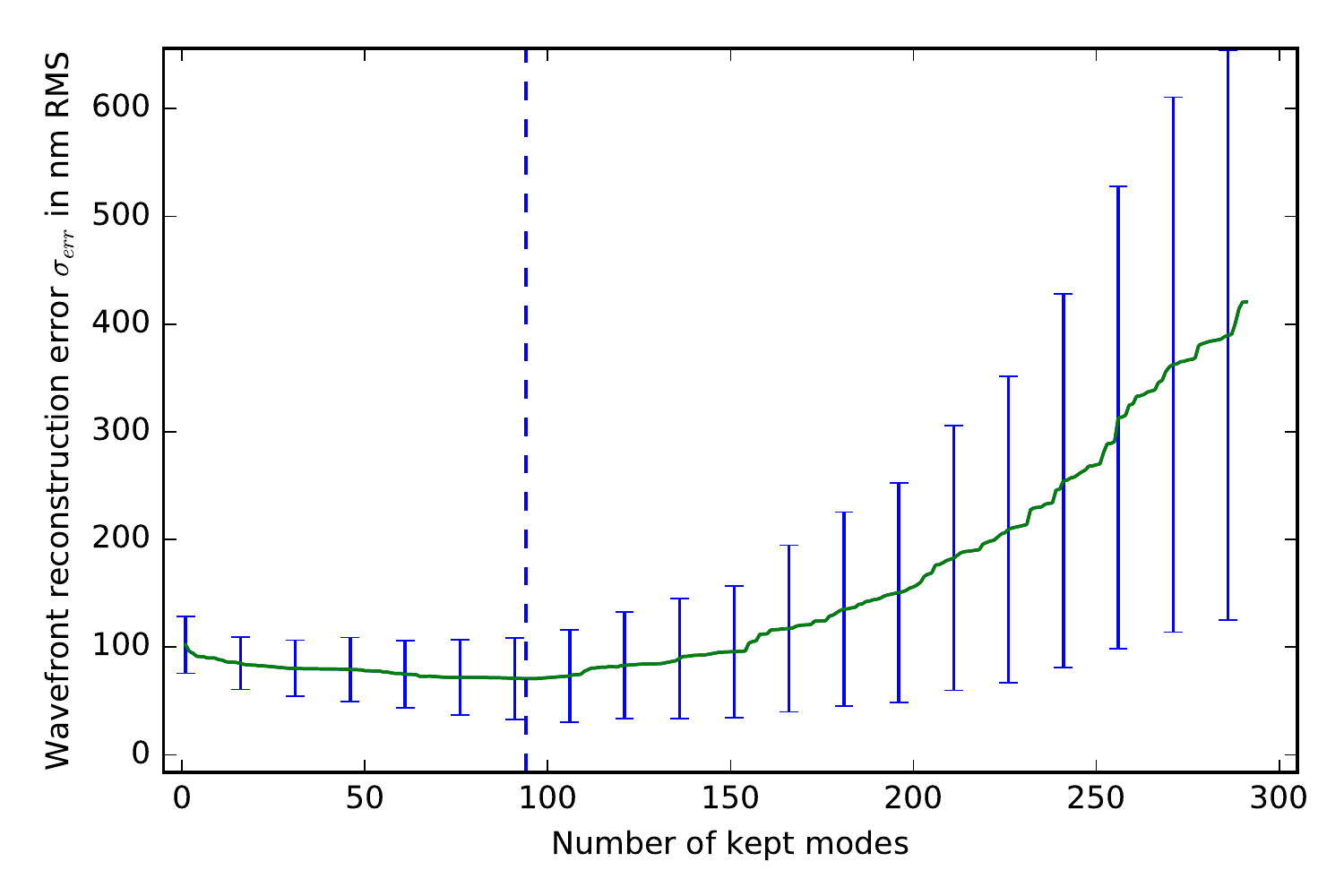}
\caption{Averaged wavefront reconstruction error $\sigma_{err}$ in nm RMS at $\lambda$=1.6\,$\mu$m as a function of the mode cutoff number. The curve and error bar represent the mean and the dispersion of the reconstruction error for 100 phase maps. The dashed vertical line represents the optimal mode number cutoff giving the minimal reconstruction error, here $k=96$.}
\label{fig:kept_modes}
\end{figure}

The reconstruction quality for the experiment can be tuned by adjusting the number of kept modes for the pseudo-inverse matrix $\mathbf{A_{k}^{+}}$. To determine the optimal mode cutoff $k$, we consider a set of 100 phase maps based on a linear combination of IE modes, with an averaged RMS wavefront error of 100$\pm$25\,nm at $\lambda=$1.6\,$\mu$m. The reconstruction error is first estimated as a function of the mode cutoff for each phase map and then, we stack the resulting errors for the whole set of maps to produce the set average and dispersion. We analyze the impact of the number of kept modes in the linear model in Figure \ref{fig:kept_modes} plot, in which two distinct regimes are observed. As we increase the number of kept modes, an improvement of the reconstruction fidelity is first observed until $k=96$. Beyond this optimal value, the phase map reconstruction degrades as we increase the cutoff for the singular modes. The reconstruction is performed for phase maps from a linear combination of IE modes and exhibits the same trend as the reconstruction for a linearly combined Zernike mode phase map \citep{Martinache2013}. In the rest of the paper and for the sake of simplicity, we set $k=100$ since this value exhibits almost the same reconstruction error as $k=96$.

The discrete model used for the pupil is identical to that of \citet{Martinache2016} in which spiders are not considered. While for Zernike modes such an aspect is not critical, these IE aberrations are strongly tied to the exact location of the spiders. A proper treatment of these spiders is beyond the scope of this paper and will be the object of a future work. As the next section will show, despite its apparent limitations, the current implementation already provides a satisfactory means to control the IE modes in closed loop.

\section{Performance on SCExAO}\label{sec:SCExAO}
\subsection{Integration on a real system}
We now use the APF-WFS with Subaru/SCExAO to demonstrate its capability to compensate for the IE on a real instrument. Our experiments are first led with the internal source and then on sky with the light coming from the Subaru Telescope combined with the adaptive optics system AO188. We briefly review the relevant features of the instrument for our tests. A more complete description of SCExAO and its capabilities can be found in \citet{Jovanovic2015}. 

SCExAO relies on several control loops to measure wavefront error contributions in different spatial frequency regimes and correct for them with a 2000-actuator DM (45 actuators across the pupil diameter). These loops involve the two instrument channels: (1) a visible channel containing the main loop with a Pyramid wavefront sensor to calibrate the atmospheric turbulence and (2) a near-infrared channel including an arm with the Lyot-based low-order wavefront sensor \citep[LLOWFS, ][]{Singh2015} and an InGaAs CMOS science camera in the final image plane for PSF acquisition. The visible channel also includes VAMPIRES, a module based on non-redundant aperture masking interferometry \citep{Tuthill2000} and differential polarimetry to image the innermost regions of protoplanetary disks \citep{Norris2015}. Its camera relies on a 512x512 pixel EMCCD detector with 16\,$\mu$m pixel size. In the near-infrared channel, the camera is based on a 320x256 pixel size detector with a 30\,$\mu$m pixel size and acquires images at a 170\,Hz frame rate with a 140\,e$^{-}$ RMS readout noise. Pupil imaging is made possible by inserting a lens in front of the camera and translating the conjugation of the camera to a pupil plane. The near-infrared channel also includes a filter wheel with an asymmetric pupil mask in a plane that is conjugated to the telescope pupil plane, enabling APF-WFS measurements with the CMOS camera. 

\begin{figure}[!ht]
\centering
\includegraphics[width = 0.5\textwidth]{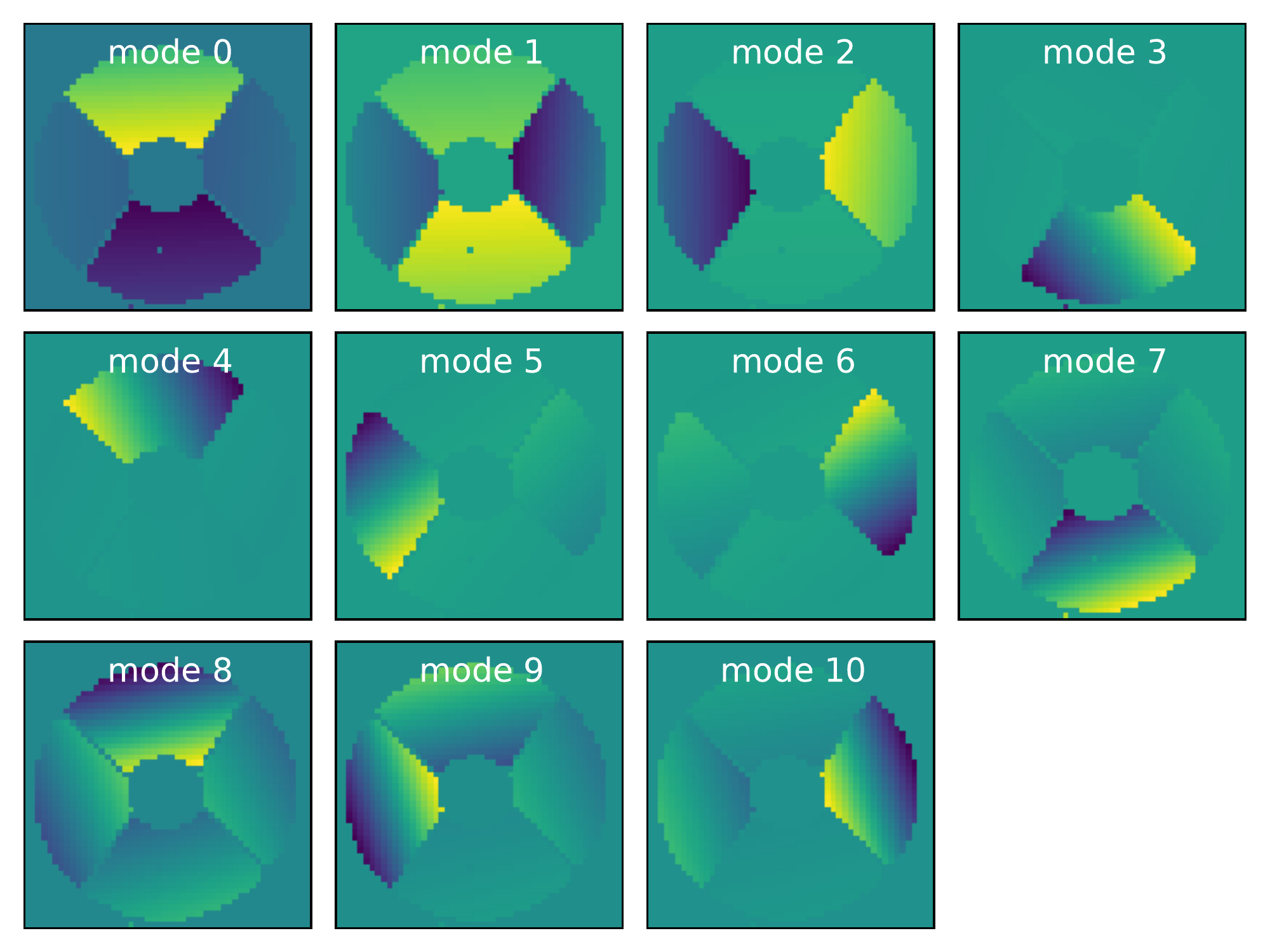}
\caption{Display of the 11 modes for the DM to control the island effect on SCExAO. These modes represent an orthogonalized version of the modes that are shown in Figure \ref{fig:11modes_DM}. They include the dead actuator of the SCExAO DM that is visible in the bottom quadrant of the pupil, for a few of these modes.}
\label{fig:11modes_DM_SCExAO}
\end{figure}

Every control loop sensor or corrector has a set of basis modes to control the corresponding device. In previous work, \citet{Martinache2016} used a Zernike basis set of modes for the APF-WFS to measure the residual low-order aberrations that are present in the system. This basis, however, does not encompass the IE modes to calibrate. In the following, we make use of an orthogonalized set of IE modes that are achieved with SVD and shown in Figure \ref{fig:11modes_DM_SCExAO} to evaluate the calibration of the differential piston effects and the simultaneous correction of mixed modes.

For our tests with the asymmetric pupil mask, we keep the pupil model with 292 subapertures from \citet{Martinache2016}. Unless otherwise stated, we keep the first 100 model singular modes for the phase reconstruction and for each experiment configuration, we acquire 100 images in $H$ band with a 10~$\mu$s integration time for the calibration step to produce the response matrix of our system. These images are then dark subtracted, recentered, and finally stacked together to produce a final averaged image. 

\subsection{In-lab results with the internal source}
We first perform tests with the internal source to validate our focal plane wavefront sensor for the IE measurements in the laboratory. We study our concept with the analysis of its response to the 11 IE modes and we operate it in closed-loop operation for the control of a given IE phase map. A control loop gain of 0.05 was used.

\subsubsection{Sensor response to IE modes}
\begin{figure}[!ht]
\centering
\includegraphics[width = 0.5\textwidth]{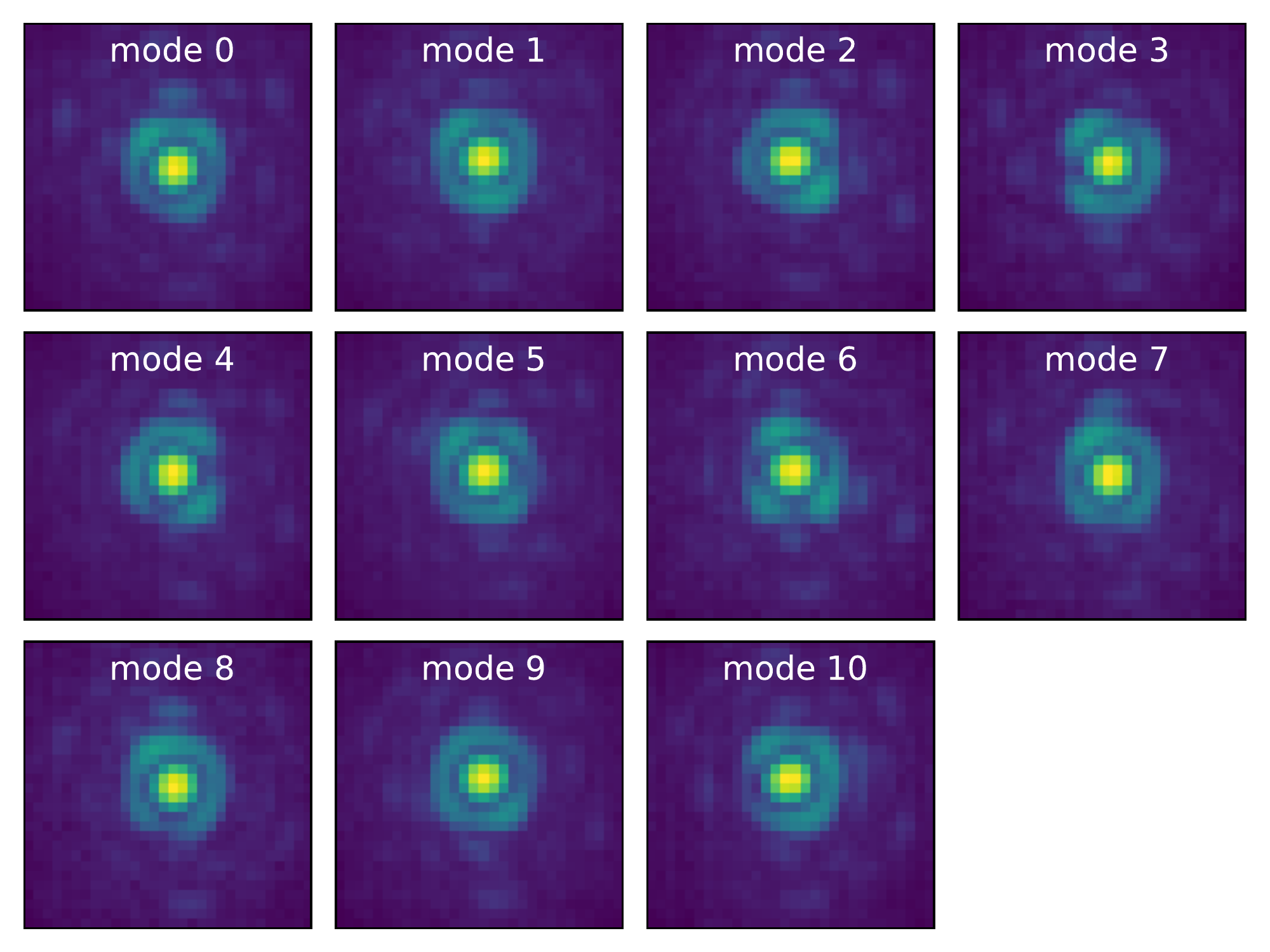}
\caption{PSFs for the 11 modes corresponding to Figure \ref{fig:11modes_DM_SCExAO} with 100\,nm RMS wavefront error, in the presence of the asymmetric pupil mask.}
\label{fig:11PSFs_SCExAO}
\end{figure}

Figure \ref{fig:11PSFs_SCExAO} shows the $H$-band PSFs resulting from the introduction of each of the 11 orthogonalized modes on the DM with 50\,nm RMS amplitude, that is, 100\,nm RMS wavefront error. Although quite similar at first glance, small differences can be seen by inspecting the first diffraction ring and the speckle field at larger angular separations. Such a set of images was later used to perform the APF-WFS modal calibration step before operating wavefront control.

\begin{figure*}[!ht]
\centering
\includegraphics[width=\textwidth]{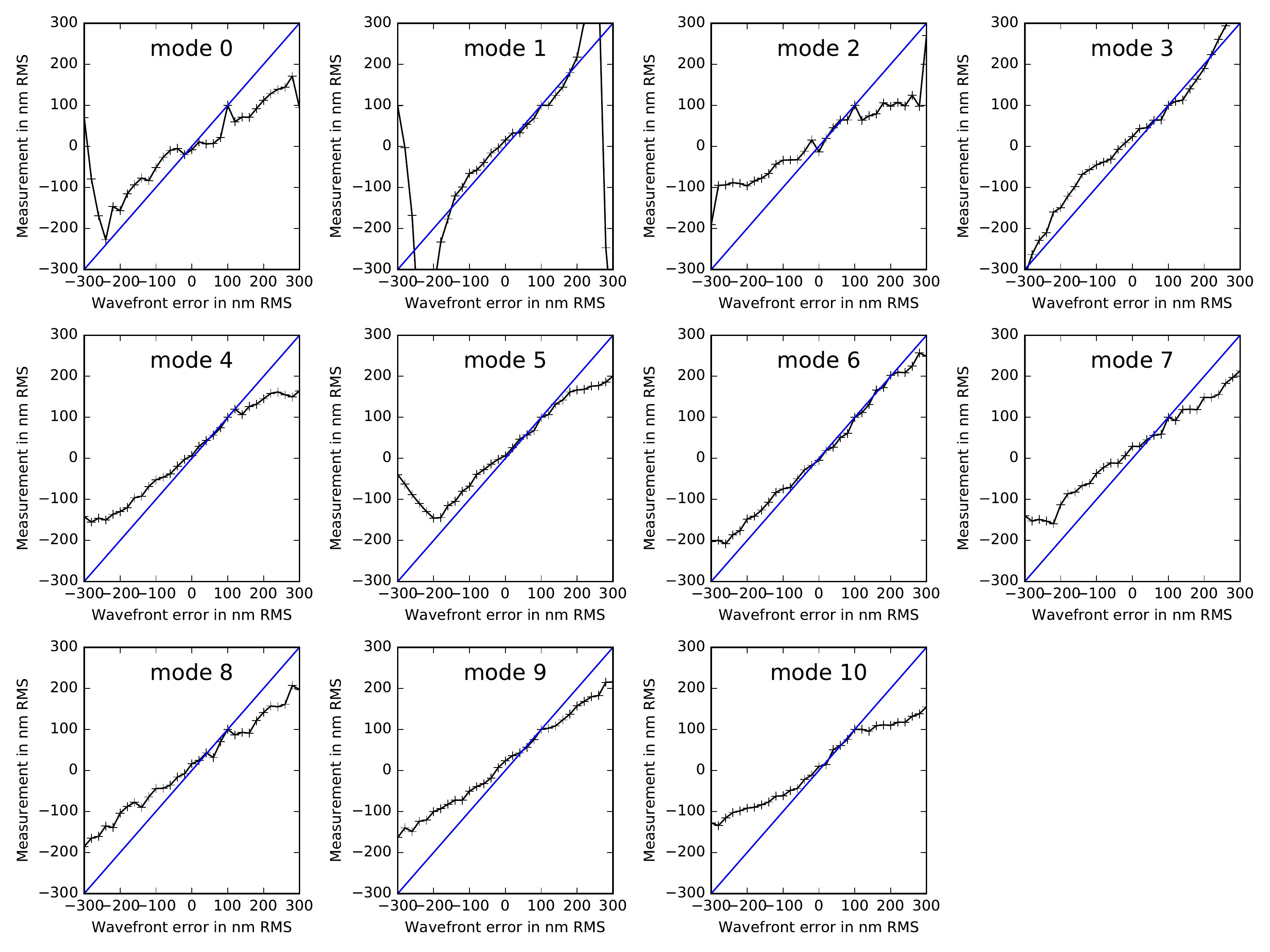}
\caption{APF-WFS response to the 11 IE modes for a pupil model with a cutoff of 100 model singular modes.}
\label{fig:lwe_lin_resp}
\end{figure*}

These images are, however, acquired at a single excitation amplitude. To provide a broader description of the sensor behavior, we analyze the sensor response to a wide range of excitations for each of the 11 modes (see Figure \ref{fig:lwe_lin_resp}). As one would expect from the small aberration regime, each IE mode has a linear response over a range of at least $\pm$~100~nm RMS wavefront error and for most of them up to $\pm$~200~nm RMS. For most of the modes, this behavior is also in good agreement with the linear response and capture range that were previously observed by \citet{Martinache2016} with APF-WFS for the excitation of the first eight Zernike modes beyond tip and tilt.

At large absolute wavefront errors, some of the response curves exhibit inflection points and non-linearities, which do not necessarily lie in the sensor itself. Instead, these errors are most likely related to two things: the pupil discretization errors that have been already discussed in Section \ref{subsec:closed-loop} and the difficulty of generating pure IE modes that introduce actual wavefront discontinuities with a continuous membrane DM. 

These tests have been performed with a phase reconstruction based on 100 kept modes for the model. We also study the evolution of the linear behavior of our concept to the modes with respect to the number of kept model modes by repeating additional tests. For a model mode cutoff larger than 70 modes, the sensor behavior shows almost no difference in terms of linearity response and capture range. All-in-all, such a linear response is promising for further closed-loop correction of the IE.

\subsubsection{Wavefront control loop}
\begin{figure}[!ht]
\centering
\includegraphics[width=0.5\textwidth]{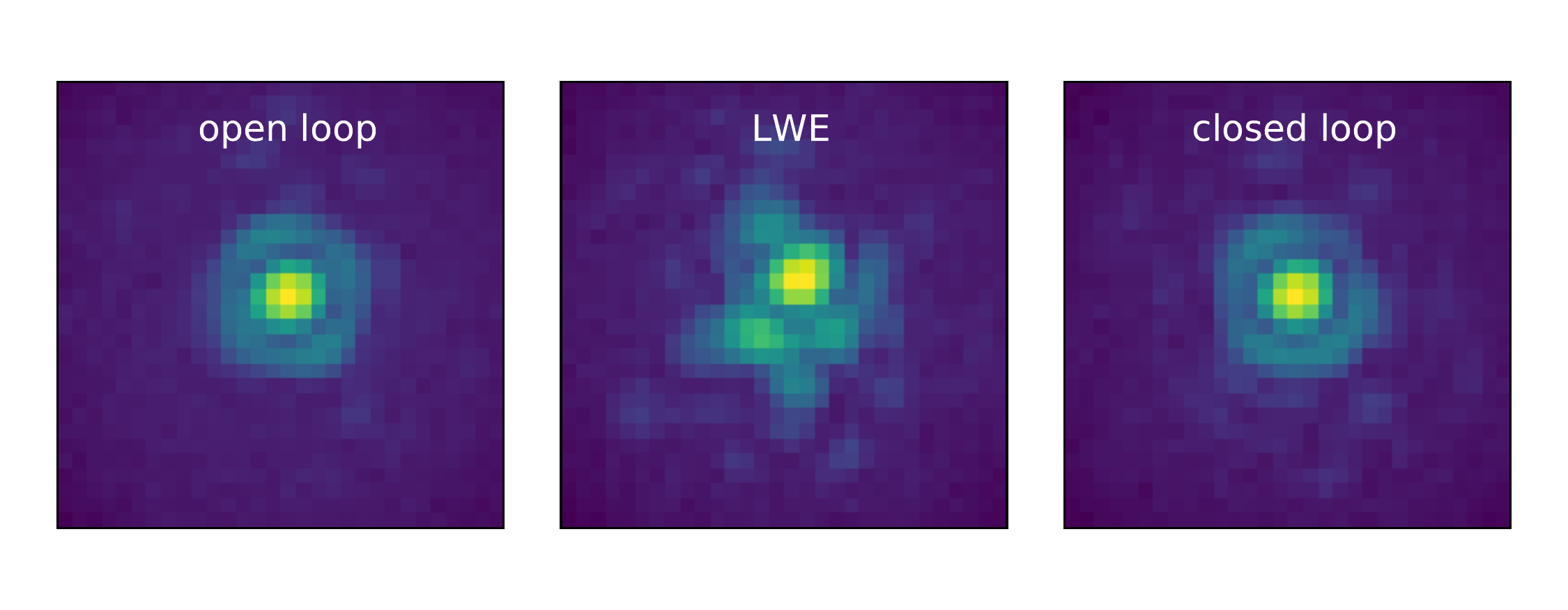}
\caption{Images at different steps of the experiment from left: in open loop without IE, after introduction of the IE on the DM, and after closed-loop wavefront control. The initial and final PSFs exhibit similar clean structures.}
\label{fig:close-loop_img}
\end{figure}

\begin{figure}[!ht]
\centering
\includegraphics[width=0.5\textwidth]{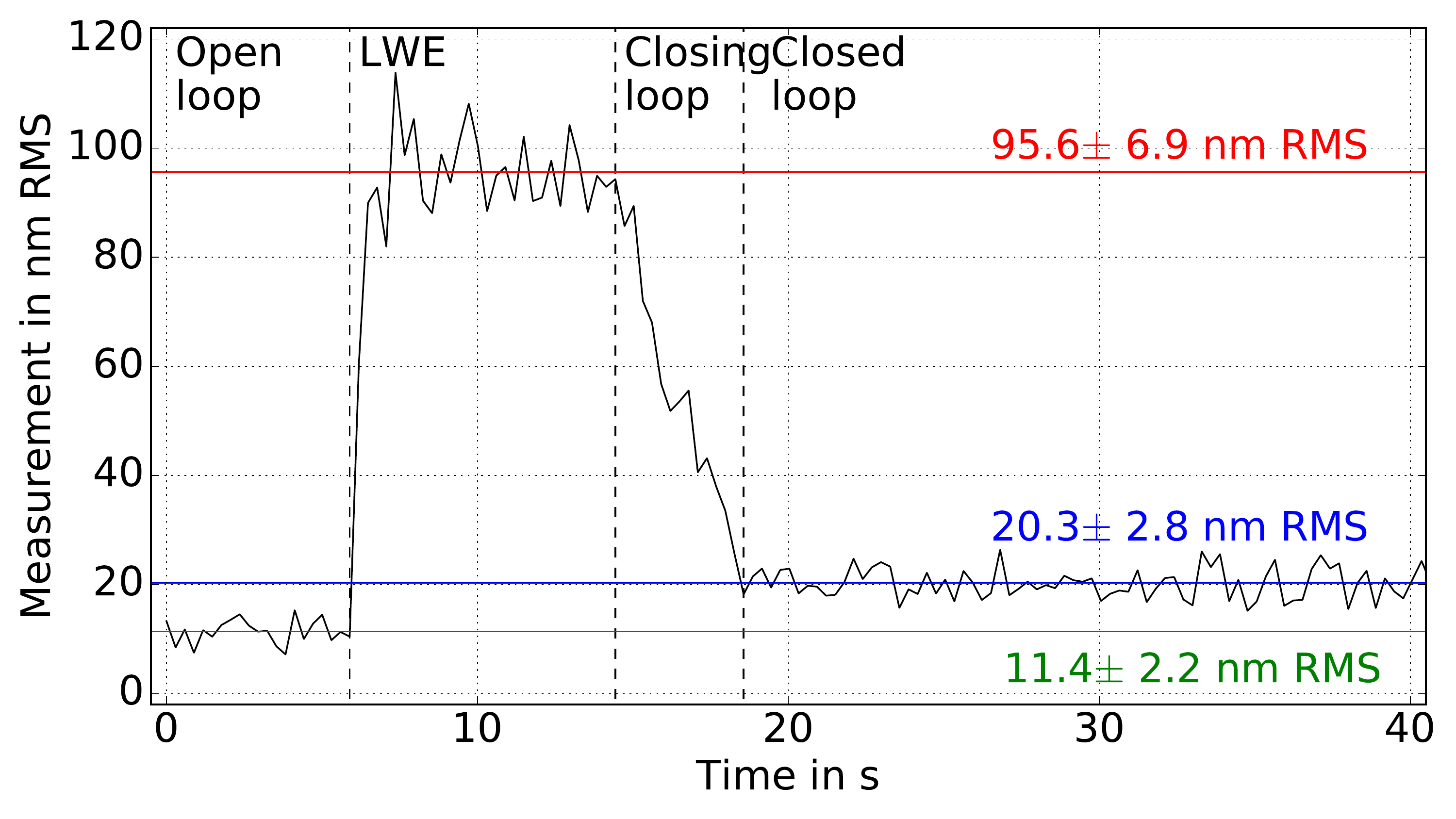}
\includegraphics[width=0.5\textwidth]{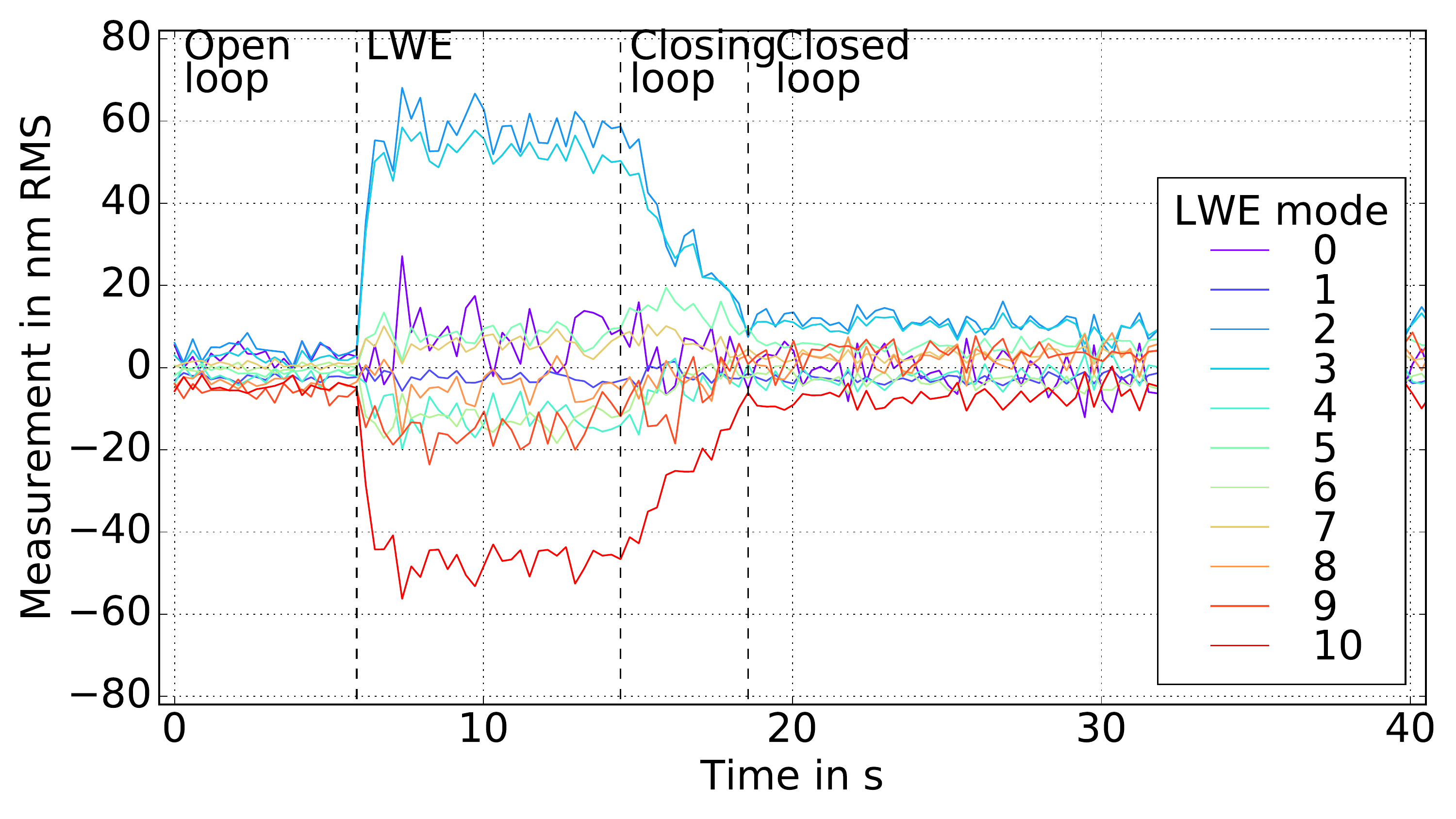}
\caption{Temporal evolution of the wavefront error during the experiment to test the control loop with the IE. The successive steps are represented by the dashed vertical lines: open loop, introduction of IE map on the DM, closed-loop operation, and stability regime after closed-loop convergence. \textbf{Top}: total wavefront error. \textbf{Bottom}: wavefront error contribution for the coefficient corresponding to each IE mode.}
\label{fig:close-loop}
\end{figure}

After the modal calibration step, we investigate closed-loop wavefront control using the APF-WFS for an arbitrary IE phase map. Figures \ref{fig:close-loop_img} and \ref{fig:close-loop} (top) show the evolution of the PSF and the total wavefront error over time. In the following, wavefront errors are estimated with their averaged value and the respective dispersion over the considered stage period. As a first step, we work with SCExAO using the DM with a shape that compensates for the aberrations from its curvature at rest, providing a clean PSF on the camera.  At this stage, the control loop is open and we estimate a total amount of 11.4 $\pm$ 2.2 nm RMS wavefront error with the APF-WFS. A few seconds later, we introduce the IE map on the DM and estimate 95.6 $\pm$ 6.9 nm RMS wavefront error while observing a degraded PSF on the camera. Another ten seconds later, we close the control loop and observe a rapid convergence towards a stable residual wavefront error of 20.3 $\pm$ 2.8 nm RMS, enabling a good quality recovery of the initial PSF.

Figure \ref{fig:close-loop} bottom plot shows the temporal evolution of the contribution for each IE mode throughout the whole experiment. The IE coefficients are obtained after projection of the corrected phase map into the IE mode basis. After introducing a given IE, our closed-loop operation clearly improves the quality of the wavefront for all the modes after just a few iterations in a matter of a few seconds.

The correction, however, does not allow us to achieve the level of wavefront errors before the introduction of the IE. The residual errors left after closing the loop find their origin in the presence of two sources of noise: the large readout noise of the camera (140\,e$^{-}$ RMS) and the photon noise as it was previously indicated by \citet{Martinache2013}. Another limiting factor is the capture range of the camera with a maximum 14-bit output format.

A more advanced IR camera will allow us to overcome the current limitations due to the detector. From the post processing point of view, we are also currently exploring an improvement in the image processing of our sensor with the introduction of a covariance matrix to whiten the noise effects. Combined with the acquisition, the computation of wavefront makes our control loop run at a frequency of $\sim$3\,Hz, allowing us to compensate for the slow-varying IE coming from LWE in the small aberration regime The computation currently represents the bottleneck for the limited speed since the code is written in python without any attempt to optimize its speed. Further code optimization will help gain speed and reach the speed of the current camera (170Hz) but the result is already promising for a demonstration. With a more advanced camera, we also expect to increase the temporal bandwidth for our wavefront control and address IE aberrations with high temporal frequency components.

In this paper, we focus on the control of the IE modes, that is, the first three Zernike modes over telescope quadrants, while \citet{Martinache2016} worked on the compensation of the Zernike modes over the full pupil beyond the third mode. In real life, modes from both families co-exist and deteriorate the quality of images together. A simultaneous control of the combined aberrations is therefore required. Modes from both families are not necessarily orthogonal between them and simply combining them may lead to an inefficient wavefront control loop. In future work, we will investigate the control of both Zernike and IE modes with an orthogonal basis of the combined sets of modes to correct for all the aberrations simultaneously and efficiently.

\subsection{First on-sky demonstration}
After validation with an internal source, we operate our closed-loop wavefront control for on-sky demonstration with an unresolved bright star, Procyon A. Our acquisitions were obtained on the Subaru Telescope during the SCExAO run on UT 2017-03-12 over a one hour time period. A 3.2 m/s wind speed and a 0.45 arcsec seeing at 0.5\,$\mu$m were noted, corresponding to good observing conditions. We tested our wavefront control loop based on APF-WFS measurements for the compensation of the IE and we evaluated the quality of the correction based on image acquisition with the IR and visible SCExAO benches: (i) in the near-infrared at 1.6\,$\mu$m ($H$ band) using the CMOS science camera and (ii) in visible light at 750\,nm with the VAMPIRES module relying on its EMCCD camera \citep{Norris2015}. As a reminder, the SCExAO control loop {with APF-WFS} works in the near infrared and still corresponds to an AO corrected beam by AO188. 

\begin{figure}[!ht]
\centering
\includegraphics[width=0.24\textwidth]{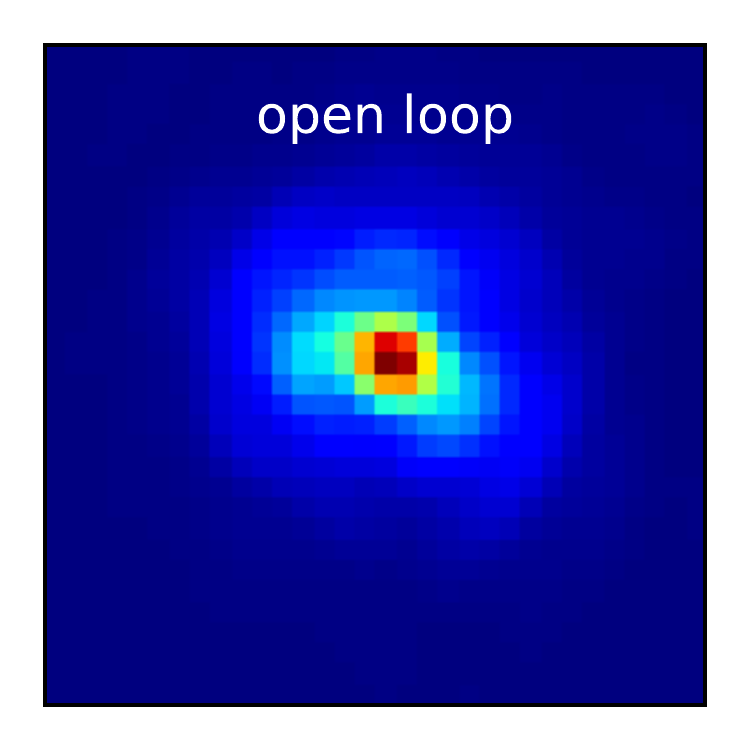}
\includegraphics[width=0.24\textwidth]{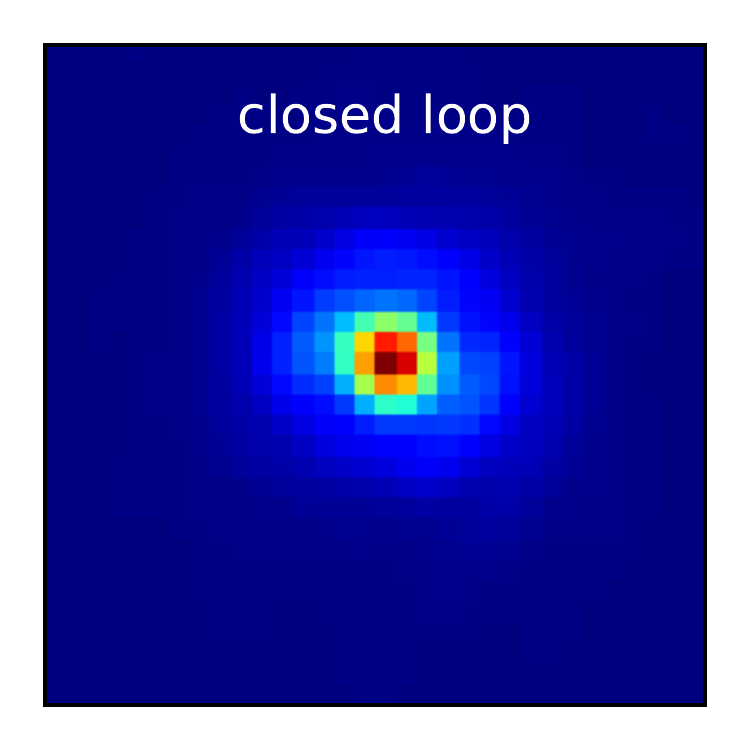}
\caption{SCExAO/VAMPIRES images of Procyon A at 750\,nm before (left) and after (right) our focal plane closed-loop wavefront control on the IE. A 0.5 power scale is used to represent these frames. Each image is normalized to its intensity peak and both frames are represented with a  different color scale to enhance the image differences between them. The star image is sharper after IE compensation.}
\label{fig:onsky_PSF_reconst}
\end{figure}

We first observe our star image in open loop in $H$ band and notice a well-structured PSF, resulting from the good and stable observing conditions during our run and the small IE amount present during our tests. Closing our control loop therefore does not show any major improvement in the PSF quality. Following our experiment in the near-infrared, we repeat the observations in the visible. As AO efficiency decreases at shorter wavelengths, there are more uncorrected aberrations in shorter wavebands, leading to an image quality in the visible that is worse than in the near-infrared. Despite this known issue, we operate our wavefront control loop and analyze our VAMPIRES images (see Figure \ref{fig:onsky_PSF_reconst}). With a frame rate of 100 images per second, we acquired 1600 short-exposure images that were dark subtracted, recentered, and finally stacked together to produce the final averaged images. Our PSFs show sharpness differences in open and closed-loop corresponding to a visual improvement in resolution. At the time of the experiment, we performed multiple switches between open and closed loop and visually observe similar image quality differences, but PSFs were only recorded for a single open and closed loop settings. Quantitatively, we estimate the Strehl ratio by measuring the PSF intensity peak in the images and find a relative increase by 37\% between the images before and after closing the loop. Such an improvement only results from the IE correction that is achieved with our wavefront control loop. Low-order non-common path aberrations not originating from IE can have a non-zero projection on the IE modes; it is therefore not possible to rule out that our control loop used IE modes to correct low-order NCPA. Further tests will require the use of an orthogonalized hybrid basis built up from both low-order Zernike and IE modes for an unambiguous demonstration. Although mostly qualitative, these first results prove encouraging. Obviously, further tests under observing conditions featuring significant IE are required to evaluate the efficiency of our algorithm for the IE in the near infrared. However, combined with the results reported using SCExAO's internal calibration source, these preliminary results already underline the ability of our algorithm based on APF-WFS measurements to compensate for the IE on-sky.

\section{Conclusion}\label{sec:conclusion}
Mostly appearing at the time of the best observing conditions, the IE results in spurious and uncontrolled aberrations that strongly limit the detection of exoplanets by direct imaging. In this paper, we present a successful calibration of the IE with a focal plane wavefront control algorithm based on APF-WFS measurements in the small aberration regime. We experimentally test our control loop with Subaru/SCExAO on the internal source and on sky. In the laboratory, we show that the loop can compensate for up to $\pm$100\,nm RMS of IE wavefront error. Preliminary testing on sky revealed a reduction of the angular resolution in the visible range, thanks to a relative improvement of the Strehl ratio by 37\%, with an online wavefront correction based on a control algorithm that operates in the near infrared. To the best of our knowledge, our demonstration represents the first real-time calibration of the IE in the context of exoplanet high-contrast observations. Such results prove extremely encouraging to ensure a maximized science return of the exoplanet imagers during favorable observing conditions. 

Our correction will further be extended to a wider capture range and a more accurate estimate of the IE wavefront errors if we manage to overcome three main limiting factors:
\begin{itemize}
\item the discretization of the pupil model to properly address the telescope struts,
\item the readout noise in the science camera to increase the accuracy of phase reconstruction with the APF-WFS algorithm,
\item the generation of accurate phase maps including discontinuities with the correction device to perform fine phase correction. 
\end{itemize}
For the first issue, we are currently working on the improvement of the discrete model to account for the telescope struts such as spiders and thus gain accuracy in the wavefront error measurements. In addition to the model refinement, we are also considering the optimization of the asymmetry to adapt the sensitivity of the measurement to the IE modes. For the last two issues, we will develop novel solutions in the next few months with numerical simulations and tests in laboratory with a new testbed related to the KERNEL project in Nice. We will explore the ultimate limits of our wavefront sensing algorithm with a fast and low readout noise InGaAs camera from FirstLight Imaging. Additionally, we will push the wavefront correction limits further by considering a segmented DM instead of a continuous facesheet device to compensate for the discontinuous phase errors from the IE. Studies on pupil segmentation with the Kernel-phase approach have already shown promising results in terms of segment cophasing \citep{Pope2014}. 

At the moment, our control loop runs at a frequency of $\sim$3\,Hz, enabling us to compensate for slow-varying IE due to low-wind conditions. A faster control loop will be explored with our fast and low readout noise camera in laboratory to increase the temporal bandwidth and further address the IE contribution with low and high temporal frequencies.

Our APF-WFS allows measurements for wavefront errors in the small aberration regime ($\varphi \ll$ 1\,rad), enabling us to partially recover the observing time that is lost for high-contrast observations of circumstellar environments due to IE. This sensor and analogous concepts based on interferometric methods including ZELDA are currently limited for the estimate of large wavefront errors. To address very large IE aberrations, we will investigate strategies based on our sensor with approaches relying on multi-wavelength measurements to increase the sensor capture range \citep[e.g.,][]{Vigan2011,Martinache2016b}.

So far, we have only considered the correction of IE modes. We are currently investigating the simultaneous control of both IE and Zernike aberration modes, relying on an orthogonal basis that encompasses both sets of modes. Such a solution will prove useful for a fast and efficient correction of the overall residual aberrations on 8\,m class telescopes with the current exoplanet high-contrast instruments but also with the future Extremely Large Telescopes (ELTs) and their instrument facilities.

The experience gained from the calibration of this unexpected effect will help the community to face unforeseen aberrations with the ELTs that exhibit a more complex architecture than the current 8\,m class telescopes and to develop wavefront control algorithms that will efficiently address unknown artifacts. In exoplanet imagery, addressing such critical aspects will rapidly bring important benefits to future observatories, with an increased capability in studying exoplanets that are fainter or closer to their host stars than the current substellar mass companions and further complete the exoplanet landscape at different age, mass, orbit, and nature.  

\begin{acknowledgements}
This work is supported by the European Research Council (ERC) through the KERNEL project grant \#683029 (PI: F. Martinache). M.N. would like to thank Alain Spang for very engaging conversations on the identification and diagnostic of the island effect by \citet{Couder1949} with a Foucault knife-edge test on an 80\,cm telescope at Observatoire de Haute Provence. This work was supported by the Astrobiology Center (ABC) of the National Institutes of Natural Sciences, Japan and the directors contingency fund at the Subaru Telescope. The authors wish to recognize and acknowledge the very significant cultural role and reverence that the summit of Maunakea has always had within the indigenous Hawaiian community. We are most fortunate to have the opportunity to conduct observations from this mountain.
\end{acknowledgements}

\bibliography{2017_mndiaye_biblio_v0}

\end{document}